\documentclass{aa} 
\def\et{et al.\ }                 %et al.%
\def\sourcea{\object{QSO\,1150+812}}
\def\sourceb{\object{BL\,1803+784}}
\def\deg{$^{\circ}$\,}

\begin{document}

\thesaurus{05 ( 05.01.1; 
             03.20.2;                   
	     11.17.4 \sourcea;
	     11.02.2 \sourceb)}
%	     11.17.4 \1150+812; 
%	     11.02.2 \1803+784)}
\title{Towards global phase-delay VLBI astrometry: Observations of 
\sourcea\ 
and 
\sourceb\
}

\author{M.A.~P\'erez-Torres\inst{1}
            \thanks{\emph{Present address:} Istituto di Radioastronomia, 
		Via P. Gobetti 101, 40129 Bologna, Italy}
        \and J.M.\ Marcaide\inst{1}
        \and J.C.\ Guirado\inst{1,2}
        \and E.\ Ros\inst{3}
        \and I.I.\ Shapiro\inst{4}
        \and M.I.\ Ratner\inst{4}
        \and E. Sard\'on\inst{5}
       }

\offprints{M.A. P\'erez-Torres} 
\mail{torres@ira.bo.cnr.it}

\institute{Departamento de Astronom\'{\i}a y Astrof\'{\i}sica, Universidad de
           Valencia, E-46100 Burjassot, Valencia, Spain
\and
  Centro de Astrobiolog\'{\i}a, INTA/CSIC, 26650 Torrejon de Ardoz, Madrid, Spain
\and
  Max-Planck-Institut f\"{u}r Radioastronomie, Auf dem H\"ugel 69, D-53121, Bonn,
  Germany
\and
 Harvard-Smithsonian Center for Astrophysics, Cambridge, MA 02138, USA
\and
 GMV, Isaac Newton 11, E-28760 Tres Cantos, 
 Madrid, Spain 
}

\date{Received 21 March 2000/ Accepted 25 May 2000}

\authorrunning{P\'erez-Torres \et}
\titlerunning{VLBI astrometry of \sourcea \, and \sourceb}

\maketitle

\begin{abstract}

On 18 November 1993, we observed 
the QSO 1150+812 and 
the BL Lac object 1803+784, nearly 15\deg apart on the sky,
with a VLBI array, simultaneously recording 
at 8.4 and 2.3 GHz. 
Using difference astrometry techniques,
we determined the coordinates of 1803+784 relative to 
those of 1150+812 to be 

\vspace*{-5pt}
\begin{eqnarray*}
% \Delta\alpha & = & \hspace{0.30cm} 6^h \, 7^m \, 33\fs 18469 \, \pm \hspace{0.07cm}\, 0\fs 00020  \\
% \Delta\delta & = &-2^\circ \, 30' \, 25\farcs 13557 \, \pm \, 0\farcs 00075 \\
 \Delta\alpha & = & \hspace{0.30cm} 6^h \, 7^m \, 33\fs 18469 \pm  0\fs 00020  \\
 \Delta\delta & = &-2^\circ \, 30' \, 25\farcs 13557 \pm 0\farcs 00075 \\
\end{eqnarray*}
\vspace*{-5pt}

\noindent
These standard errors contain estimated contributions from
the propagation medium, the effects of source opacity, 
and the possible misidentification of the reference point in 
some of the maps. 
We obtained comparably accurate and consistent relative
positions of the radio sources with GPS-based ionosphere corrections,
thus demonstrating that dual-frequency observations are not required
for state-of-the-art accuracy in
VLBI astrometry. 
These results also demonstrate the feasibility of using 
single-frequency measurements for sources 
separated by 15$^\circ$ on the sky, and open the avenue for
the application of this technique on a full-sky scale.

\vspace*{0.3cm}
\noindent
\end{abstract}

\section{Introduction}
\label{sec:intro}

The establishment of a radio reference frame 
with submilliarcsecond accuracy 
has been a major goal of astrometrists for the last several decades. 
Centimeter-wavelength very-long-baseline interferometry (VLBI) 
group-delay astrometry of extragalactic radio sources 
routinely provides precisions at the milliarcsecond (mas) level, 
thus allowing a celestial reference frame to 
be built with corresponding accuracy (e.g., Ma et al.~\cite{ma98}).
The use of phase-delay difference astrometry (see, e.g., Shapiro et
al.~\cite{shapiro}) allows us to determine angular separations 
between pairs of radio sources at the submilliarcsecond level. 
One important advantage of phase-delay astrometry is that we 
can identify with sufficient accuracy suitable
reference points within the structures
of the radio sources  whose relative positions
we wish to determine, whereas in group-delay astrometry 
suitable reference points are not so easily identified from epoch to epoch with 
the desired accuracy.

With the phase-delay technique, 
the differences between phase delays for two radio sources are used 
to determine their relative positions, 
because these differences have reduced sensitivity  to unmodeled effects.
For sources nearby to one another on the sky, 
this technique yields statistical standard errors of only 
a few microarcseconds ($\mu$as), as in the case 
of quasars 1038+528 A and B (Marcaide \& Shapiro \cite{jmm83}; 
Marcaide et al. \cite{jmm94}),
whose components are separated by only 33$\farcs$
However, the overall standard errors are dominated by 
inaccuracies in the reference-point identification 
(e.g., Rioja et al. \cite{rioja97}). 
For sources with larger separations, 
the main contributions to the astrometric standard error 
in the relative position of two sources comes from uncertainties in the  
coordinates of the source chosen as the reference, in the value used for 
UT1--UTC, and in the effects of the propagation medium. 
For radio sources 
with separations ranging from $0.5 \degr$ to $7\degr$ on  
the sky, standard errors in relative positions of about 0.1--0.3 mas 
have been obtained
(Bartel et al. \cite{bartel86}; Guirado et al. \cite{gui95a},b,  
\cite{gui98}; Lara et al. \cite{lara96}; 
Ros et al. \cite{ros99}).

A limiting factor in centimeter-wavelength VLBI astrometry 
is the uncertainty in the contribution of the ionosphere 
to the astrometric observables, even though its  
dispersive character makes this contribution 
scale as $\nu^{-2}$ and, in principle,  
allows it to be determined accurately
(see, e.g., Thompson et al.~\cite{tms}). 
One strategy to take advantage of this scaling is to make 
simultaneous VLBI measurements in two frequency bands. 
The main disadvantage of this option is that the (fixed) 
bandwidth of the recording 
equipment has to be split between two 
frequency bands, decreasing the signal-to-noise-ratio (SNR) for each. 
Alternatively, one may compute corrections based on 
estimates of the ionosphere 
total electron content (TEC) 
obtained independently from 
Faraday-rotation measurements and, more recently, from 
the Global Positioning System (GPS).
The advantage of the latter approach 
(e.g., Guirado et al.  \cite{gui95b}, Ros et al. \cite{ros2000}), 
is that only single-band VLBI observations are needed, 
avoiding the loss of sensitivity mentioned above.

In this paper, we study 
the applicability of the phase-difference technique to 
the strong radio sources 1150+812 and 1803+784, 
separated on the sky by almost 15\degr. 
We show that reliable phase connection 
is feasible at such a large angular separation, 
and estimate the relative position
of the two sources with submilliarcsecond accuracy.
We compare the estimates of the
relative angular separation that result from use of two different methods
of removing the ionosphere contribution based 
on two different types of data, namely, phase delays from 
dual-frequency-band VLBI measurements and TEC values from GPS measurements.
The estimates from the two methods agree to within the standard 
errors from each method, 
showing that single-frequency astrometric VLBI experiments can be 
confidently carried out. 

In Sect.~\ref{sec:observations}, we briefly describe the observations; 
in Sect.~\ref{sec:sources} we describe the radio sources, and 
in Sect.~\ref{sec:reduction} the data reduction process. 
In Sect.~\ref{sec:relpos} we discuss our estimate of the relative position of the 
source 1803+784 with respect to 1150+812, and
in Sect.~\ref{sec:sens} we carry out a sensitivity analysis 
of the estimated position of 1803+784 to errors 
in other model parameters. 
Finally, in Sect.~\ref{sec:results} we summarize our main results and 
discuss their implications.

\begin{figure}[htbp !]
\vspace*{360pt}
\includegraphics{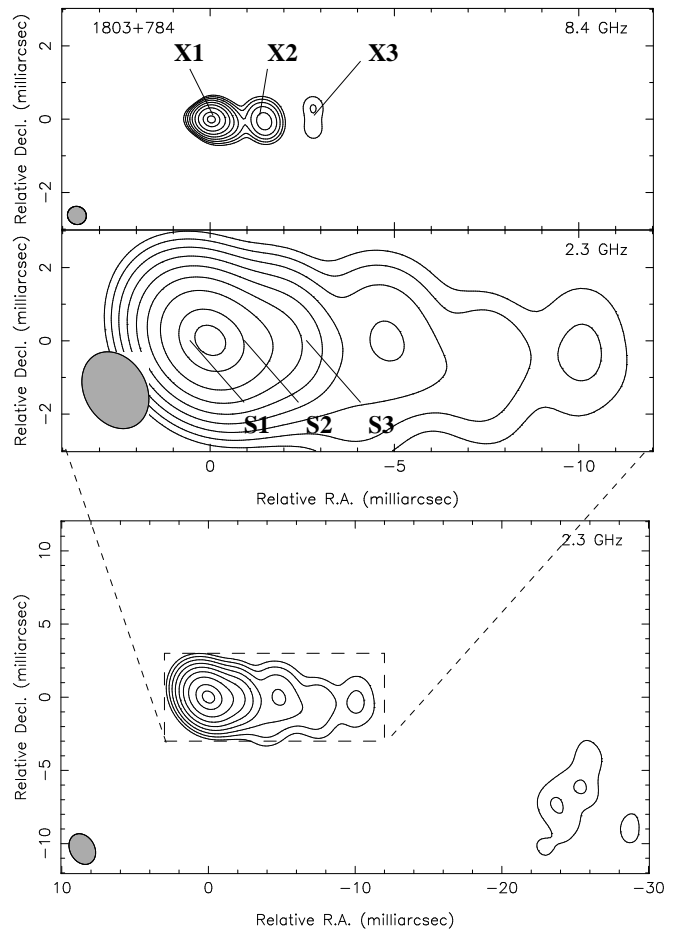}
\caption[]{
%\begin{scriptsize}
Hybrid maps of 1803+784 at 8.4 (top) and 2.3 GHz (middle and bottom).
Contours are 0.5,1,2,4,8,16,32,64, and 90\% of the peaks of
brightness, which are 1.48 Jy/beam and 1.35 Jy/beam for 8.4 and 2.3 GHz, 
respectively. 
The sizes of the restoring beams, shown shaded at the bottom left corners, are 
0.52 $\times$ 0.49 mas (PA=61$^\circ$) at 8.4 GHz and  
2.21 $\times$ 1.68 mas (PA=29$^\circ$) at 2.3 GHz.
East is left, and north is up. 
The identification of the components in 
the 2.3 GHz map is based on an overresolved map 
(not shown here). 
%\end{scriptsize}
}
\label{fig:1803}
\end{figure}

\section{Observations}
\label{sec:observations}

We observed the radio sources 1150+812 and 1803+784 
on 18 November 1993 for 12 hours, in right circular polarization, 
simultaneously recording at two frequency bands 
(X-band $\approx$ 8.4\,GHz and S-band $\approx$ 2.3\,GHz). 
We used the following radio telescopes 
(in parenthesis we give the site 
symbol used in this paper, the diameter, and the 
location of the telecope): 
Effelsberg (B, 100\,m, Germany); 
Medicina (L, 32\,m, Italy);
Onsala (S, 20\,m, Sweden); 
Fort Davis (F, 25\,m, Texas); 
Hancock (H, 25\,m, New Hampshire);
North Liberty (I, 25\,m, Iowa);
Owens Valley (O, 25\,m, California);
Los Alamos (X, 25\,m, New Mexico);
and the phased VLA (Y, 130\,m equivalent, New Mexico). 
The European antennas (B, L, and S) recorded in Mark III mode A, 
covering a total bandwidth of 56~MHz, 
via seven adjacent channels 
spanning 
8,403 to 8,431~MHz and seven such channels 
spanning 2,273 to 2,301~MHz. 
The American antennas used the Mark III recording system
in mode B, covering a total bandwidth of 28~MHz, with four 
contiguous channels spanning 8,403 to 8,419~MHz and three 
such channels spanning 2,289 to 2,301~MHz. 
We used an observing cycle consisting of 
2 min  observing 1150+812,
1.5 min antenna slewing, 2 min observing 1803+784, 
and 1.5 min antenna slewing back to 1150+812, making a total cycle
duration of 7 min. 

\begin{figure*}[htbp]
%\begin{figure*}[p]
%\vspace*{70mm}
\vspace*{85mm}
\includegraphics{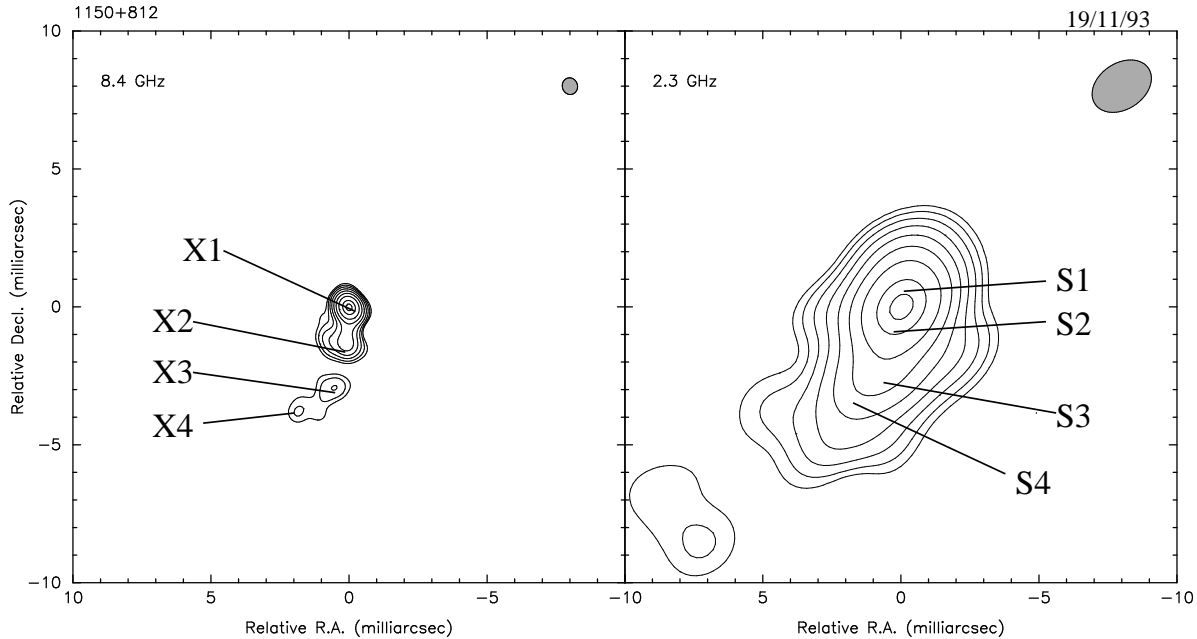}
\caption[]{
%\begin{scriptsize}
Hybrid maps of 1150+812 at 8.4 (left)  and 2.3 GHz (right).
Contours are 0.5,1,2,4,8,16,32,64 and 90\% of the peaks of
brightness, which are 0.90 Jy/beam and  0.74 Jy/beam for 8.4 
and 2.3 GHz, respectively. 
The sizes of the restoring beams, shown shaded at the upper right corners,
are 0.61 $\times$ 0.55 mas (PA=15$^\circ$) 
at 8.4 GHz and 2.33 $\times$ 1.67 mas (PA=-56$^\circ$) at 2.3 GHz.
The identification of the components in 
the 2.3 GHz map is based on an overresolved map 
(not shown here; see text). 
%\end{scriptsize}
}
\label{fig:1150}
\end{figure*}

%For the baselines involving stations recording in different modes, 
%only the overlapping frequency channels were correlated.
The data were correlated at the MPIfR correlator in Bonn, Germany.
Since both sources were strong ($>$ 1 Jy) at the epoch of observation,
we detected each of them with
high SNR within the 2 min integration time,
at both frequencies and for each baseline, except those baselines 
involving the Onsala antenna. 
After correlation, we exported the VLBI observables
(fringe amplitude and phase, group delay, and phase-delay rate)
for the reference frequencies $\nu_{\rm X}$ = 8413~MHz 
and $\nu_{\rm S}$ = 2295~MHz,  and  calibrated the fringe amplitudes 
using the information provided by the staffs at the 
observing antennas.
In our astrometric analysis, we discarded the data from 
Onsala due to the low SNR of the detections with this antenna; and 
from the phased VLA, since this antenna recorded only at 8.4 GHz and 
was used solely to increase the SNR of our hybrid maps. 
In addition, we discarded data from Effelsberg for the interval
18:30 until 23:28, since this antenna, too, only observed at 8.4 GHz during 
that interval.
To obtain our hybrid maps, we
self-calibrated, Fourier inverted, and {\sc CLEAN}ed the visibility data
using the Caltech program DIFMAP 
(Shepherd et al. \cite{shepherd}).

\section{The Radio Sources}
\label{sec:sources}

%Maps of the radio sources 1803+784 and 1150+812
%have been previously shown  by Eckart et al. (\cite{eckart87}) 
%and Witzel et al. (\cite{witzel88}).
The radio source 1803+784 has been optically identified 
with a BL Lac object at z=0.684 (Stickel et al.~\cite{stickel91}). 
%implying a luminosity distance of about 3.6 Gpc 
%($H_0$ = 65 km s$^{-1} \,$ Mpc$^{-1}$, $q_0$ = 0.5). 
This source has a nearly flat spectrum from millimeter to decimeter 
wavelengths (Strom \& Biermann \cite{strom91}) and
displays many interesting properties 
(Biermann et al. \cite{biermann81}), 
including rapid variability, compactness, and strong X-ray emission.  

Our 8.4 and 2.3 GHz maps of 1803+784 (Fig.~\ref{fig:1803}) 
show a jet pointing westward.
The hybrid map at 8.4 GHz shows the jet extending about 3 mas,
with three components, labeled X1, X2, and X3.
Most of the emission comes from 
components X1, identified as the core, and X2, about 1.5 mas west of X1. 
Component X3 lies about 2.8 mas west of X1 and is much fainter than X1 and X2.
Our map at 2.3 GHz also shows the jet, which, as in  
the maps of Ros et al. (\cite{ros99}) and Eckart et al. (\cite{eckart87}), 
extends up to about 25 mas and is slightly bent towards  
the southwest. 
At 2.3 GHz, most of the emission is concentrated within 3 mas of the peak,
but emission is also discernible at points about 5, 8, and 10 mas west of S1
(Fig.~\ref{fig:1803}, middle). 
These latter bright regions, especially the innermost 
and the outermost, may correspond to real features of the jet, 
but like the questionable features farther out, 
they are not relevant for our
astrometric purposes and, therefore, are not labeled.

The radio source 1150+812 is a QSO 
%of visual magnitude \mv $ = 18.5$ 
with redshift  $z=1.25$ (Hewitt \& Burbidge \cite{hb93}). 
%which corresponds to a luminosity distance  of $\sim 6.9$ Gpc. 
For this source, our 8.4 and 2.3 GHz maps (Fig.~\ref{fig:1150})
indicate a curved jet directed to the south and west.
The 8.4 GHz map contains  at least four components, 
labeled X1 through X4, from north to south.
Most of the emission 
comes from the components X1 and X2, 
which are separated by  $\sim 1.5$ mas. 
Extended emission is still discernible at a distance of $\sim 4$ mas from 
X1 (components X3 and X4). 
At 2.3 GHz, the emission is concentrated within the innermost 5 mas,
but emission is also evident farther to the southeast. 

Our maps of 1803+784 and 1150+812 are in good agreement with maps 
of similar resolution from 1995 
published by Fey \& Charlot (\cite{fc97}). 

\section{Astrometric Data Reduction}
\label{sec:reduction}

Our strategy in the astrometric data analysis was to obtain a set 
of ``connected'' phase delays that could be analyzed via
weighted-least-squares to estimate the relative 
position of the two sources. 

%%%%%%%%%%%%%%%%%%%%%%%%%%%%%%%%%%%%%%%%%%%%%%%%%%%%%%%%%%%%%%%%%%%%%%%%%%%
\renewcommand{\baselinestretch}{0.5} 
%
%___________________________________________  TABLE WITH THEORETICAL MODEL
%
%\begin{table*}[h!]
\begin{table*}[htbp!]
%\begin{table*}[p]
%\tighten
\caption{A priori values of the parameters of the theoretical
model used in the estimates of the source positions\label{table:model}}
\label{tab:model}
\begin{tabular}{lllllll} 
%\[
\hline
\multicolumn{7}{l}{\bf Source coordinates (J2000.0)$^{\rm a}$} \\
1150+812 & $\alpha = 11^{h} 53^{m} 12\rlap{.}^s 499188$ & $\delta = 80^{\circ} 58' 29\rlap{.}''15453$ & & & & \\
1803+784$^{\rm b}$ & $\alpha = 18^{h} 00^{m} 45\rlap{.}^s 683922$ & $\delta = 78^{\circ} 28' 04\rlap{.}''01852$ & & & & \\
\end{tabular}
%\]
\[
\begin{tabular}{llrrrrcc}
\multicolumn{7}{l}{\bf Antenna site coordinates and atmosphere parameters} \\
%  &              &              &             & 
%						       & \multicolumn{2}{c}{Mean} \\
\multicolumn{2}{l}{\small Radio}
  &\multicolumn{3}{c} {\small Cartesian coordinates [m]} & \multicolumn{1}{c}{\small Axis} 
						       & \multicolumn{2}{c}{\small Mean atmosphere} \\
\multicolumn{2}{l}{\small telescopes$^{\rm a}$} 
  &    &         &         & \multicolumn{1}{c}{\small offset \,[m]} 
						       & \multicolumn{2}{c}{\small zenith delay$^{\rm c}$ [ns]}\\
  &     & \multicolumn{1}{c}{X} 
		 & \multicolumn{1}{c}{Y} 
				& \multicolumn{1}{c}{Z} 
					      &        &  $\tau_{\rm dry}$ 
 							     & $\tau_{\rm wet}$\\
B & Effelsberg    &  4033947.51   &   486990.46  & 4900430.75  & 0.00 & 7.5 & 0.1 \\
F & Fort Davis    &  --1324009.08 & --5332181.98  & 3231962.49  & 2.12 & 6.4 & 0.1 \\
I & North Liberty &  --130872.19  & --4762317.13  & 4226851.05  & 2.13 & 7.5 & 0.1 \\
L & Medicina      &  4461370.05   &   919596.77  & 4449559.16  & 1.83 & 7.7 & 0.2 \\
O & Owens Valley  &  --2409150.05 & --4478573.27  & 3838617.42  & 2.13 & 6.8 & 0.1 \\
X & Los Alamos    &  --1449752.31 & --4975298.60  & 3709123.95  & 2.14 & 6.2 & 0.1 \\
   
\end{tabular}
\]
\[
\begin{tabular}{lll}
{\bf Earth Tides}  & \\
%Radial Love number, {\bf h} = 0.60967 &  Horizontal Love number,  {\bf l} = 0.085 &
%Love numbers$^{\rm e}$: {\bf h} = 0.609, {\bf l} = 0.0852 & 
Radial Love number$^{\rm d}$, {\bf h} = 0.60967 &
Tidal lag angle, $\theta$ = 0$^\circ$.0 \\[1mm]
Horizontal Love number$^{\rm d}$, {\bf l} = 0.085 & \\  \\
%  {\bf Earth Orientation Parameters$^{\rm a,d}$} &  & \\
    {\bf Precession Constant (J2000.0)$^{\rm e}$ } 
  & {\bf Polar Motion and UT1} 
  & {\bf Nutation}  \\
%  & Wobble: {\bf x}=$-0\rlap{.}''04594$ {\bf y}=$0\rlap{.}''44273$ 
    {\bf p}=5029$''$.0966/Julian Century 
  & Interpolated from daily values 
  & Interpolation of IAU 1980 Nutation Model  \\
%    Corrections from IERS for JD=2449310  
    {\bf Mean obliquity (J2000.0)$^{\rm e}$}  
%  & UT1$-$UTC =0.299869 ${\rm s}$   
  &  from IERS (\cite{iers})
  & with IERS (\cite{iers}) daily corrections \\
%    $d\psi=-0\rlap{.}''01977 ~~~d\epsilon=-0\rlap{.}''00470$ 
    $\epsilon_0$=23$^\circ26'21\rlap{.}''$448=84381$\rlap{.}''$448    
  &
  &  \\
\hline 
\end{tabular}
\]
%\clearpage
\begin{list}{}{}
%\begin{scriptsize}
\item[$^{\rm a}$] {
We took 0.058\,ms and 0.140\,mas to be the standard errors in 
the right ascension and declination, respectively, of 1150+812, twice
the quoted IERS (\cite{iers}) errors. 
We also used as standard errors 
2 cm in each coordinate of every antenna location 
(conservatively
chosen to be at least twofold larger than the errors given by IERS),
0.7\,mas in each pole coordinate, and 0.04\,ms in UT1--UTC.
We calculated the location of each antenna by linear extrapolation, 
using the velocity given by the IERS, and applied to the 1993.0 position,
also given by the IERS. 
}
\item[$^{\rm b}$]
{
The coordinates of 1803+784 serve only as an initial estimate, 
and do not affect our results.
}
\item[$^{\rm c}$]{
Mean values of sampled zenith delays, computed using the 
meteorological values provided by the staff at the
observing antennas. 
}
%\item[$^{\rm d}$]
%{Note that the IERS values for nutation, polar motion, and UT1--UTC
%have a ``formal'' precision larger than their accuracy.
%} 
\item[$^{\rm d}$]
{Dahlen (\cite{dahlen}).
} 
\item[$^{\rm e}$]
{Lieske \et (\cite{lieske}).} 

\end{list}
\end{table*}
\renewcommand{\baselinestretch}{1.0} 
%%%%%%%%%% END OF TABLE WITH THEORETICAL MODEL
The observables in our experiment, after correlating and 
fringe-fitting the data, were the interferometric phase, $\phi$, 
the group delay, $\tau_G$, and the phase-delay rate, $\dot{\tau}_\phi$. 
We based our analysis primarily on $\phi$, converted first to 
the phase delay, $\tau_\phi$, 
via removal of the ``$2\pi\,n$'' ambiguities (where $n$ is an integer). 
The delay equivalent of $2\,\pi$ in phase is  $\sim 120$ ps at 8.4 GHz 
and $\sim 440$ ps at 2.3 GHz.
To remove these ambiguities we used a ``phase connection'' technique
(see, e.g., Shapiro et al. \cite{shapiro}) that can be outlined as follows: 
We constructed a model of the phase delays, based on 
a model of the geometry of our interferometric array 
and the radio sources (Table~\ref{tab:model});  the propagation medium;
and the clock behavior at each station relative to a reference.
We estimated the parameters of this model 
via weighted-least-squares analysis of the observed phase-delay 
rates and group delays.
We used an improved version of the program VLBI3 
(Robertson~\cite{robertson}) to carry out this weighted-least-squares 
analysis. 
The resultant model of the phase delays was accurate enough to 
allow phase connection,
i.e., the elimination of the $2\pi\,n$ ambiguities for almost all
of the observations (see below), via a suitable iterative scheme that 
took advantage of the closure relations over triangles of baselines. 
Since we phase-connected the data independently for each source, we 
used weighted-least-squares to verify -- successfully -- the 
consistency of the ``overall resolution of ambiguities'' 
of the delays for the two sources (see Ros et al.~\cite{ros99} for
a detailed discussion of this step).
During our analysis, 
we discarded all data from  
the Hancock antenna, since our phase connection failed
at both frequencies.
We also discarded the data from the Medicina
antenna from 18:30 to 23:48, since 
in this time period our phase-connection appeared to be unreliable.  
(These failures are likely the result of some
combination of tropospheric, ionospheric, and instrumental
effects.)

\subsection{Source-Structure Contribution}
\label{sec:structure}

The extended structure of even a ``compact'' radio source 
often makes a significant contribution to the phase delays. 
Such a contribution depends on the point chosen as a reference on the map.
Identifying a reference point in a reliably epoch-independent manner 
is crucial for the use of our method to compare positions obtained 
from different epochs. 
For each source and frequency band, 
our procedure was to choose the peak of brightness (POB) 
as the reference point.  
To find the POB, we constructed fine-grained maps
of the two radio sources
using pixel size 0.01 mas at 8.4 GHz and 0.03 mas at 2.3 GHz. 
We identified the brightest pixel in each such map as the POB 
and defined a new coordinate system with 
that reference point as its origin. 
We then computed from these maps the structure-phase contribution 
to each phase delay. 
These contributions -- up to $\sim$ 25 ps at 8.4 GHz and 
$\sim$ 55 ps at 2.3 GHz -- 
were removed from the phase delays, to effect a point-like source
at each POB.
The standard deviations assigned to each POB location included 
uncertainties due to (a) the pixel size used for each map, 
which was about  0.01 (0.03) mas 
at 8.4 (2.3) GHz, for each source; and 
(b) SNR for the peak, which was 
about 0.04 (0.27) mas at 8.4 (2.3) GHz, for each source. 
Note, however, that when the delays at the two frequencies are combined 
to form ``plasma-free'' delays, errors in the phase delays 
at 2.3 GHz are scaled down 
by a factor ${[(\nu_X/\nu_S)^2 - 1]}^{-1} \approx 0.08$
(see Sect.~\ref{sec:opacity}), and therefore
the uncertainty of the POB at 2.3 GHz is less significant than 
that of the POB at 8.4 GHz.
The root-sum-squares of these contributions, at each frequency, 
are indicated as the first (8.4 GHz) and second (2.3 GHz) entries 
in Table~\ref{tab:errors}, 
and are dominated by the relatively low SNR of the hybrid maps.

\subsection{Neutral Atmosphere Contribution}
\label{sec:troposphere}

The neutral atmosphere (primarily the troposphere) adds an extra delay to the 
incoming radio waves, the equivalent of up to a few meters in pathlength.
We monitored the pressure, relative humidity, and temperature 
at each observing antenna to track the 
atmosphere behavior during the observations. 
We used  the model by Saastamoinen \cite{saasta}, in which 
the atmosphere is separated into two components:
a dry component and a wet component 
(due to the water vapor in the atmosphere).
Based on this model, we
calculated a priori values for the delays for the wet and dry 
atmosphere components in the zenith direction 
for each antenna site (Table~\ref{tab:model}), 
and then used the Chao (\cite{chao}) mapping function 
to determine delays at other elevations. 
This mapping function agrees to within about 1 cm with ray-tracing
computations (Davis et al. \cite{davis85}) for antenna elevations larger than 
20$^\circ$; all of our observations  had elevation angles above this limit.
To specify adjustments to the combined (dry and wet) 
atmosphere delays during the observations, 
we used a piecewise-linear function
characterized by zenith delay values at epochs about three hours apart. 
Errors in the combined neutral-atmosphere delays -- mainly due to the 
wet troposphere -- are likely to be up to 0.1 ns/hr (D. Lebach, priv. comm.). 
Since the wet atmosphere zenith delay fluctuates approximately as  a random walk
(Treuhaft \& Lanyi~\cite{wet}), 
an error of 0.1 ns/hr transforms into  
a standard error of $\sqrt{3} \times 0.1 \approx 0.17$ ns every three hours. 
Therefore, in our sensitivity analysis, we allowed the atmosphere parameters at each
antenna location to vary with this standard error.

\subsection{Ionosphere Contribution}
\label{sec:ionosphere}

\begin{figure*}[htbp !]
%\picplace{5cm}
%\vspace*{100mm}
\vspace*{270pt}
\includegraphics{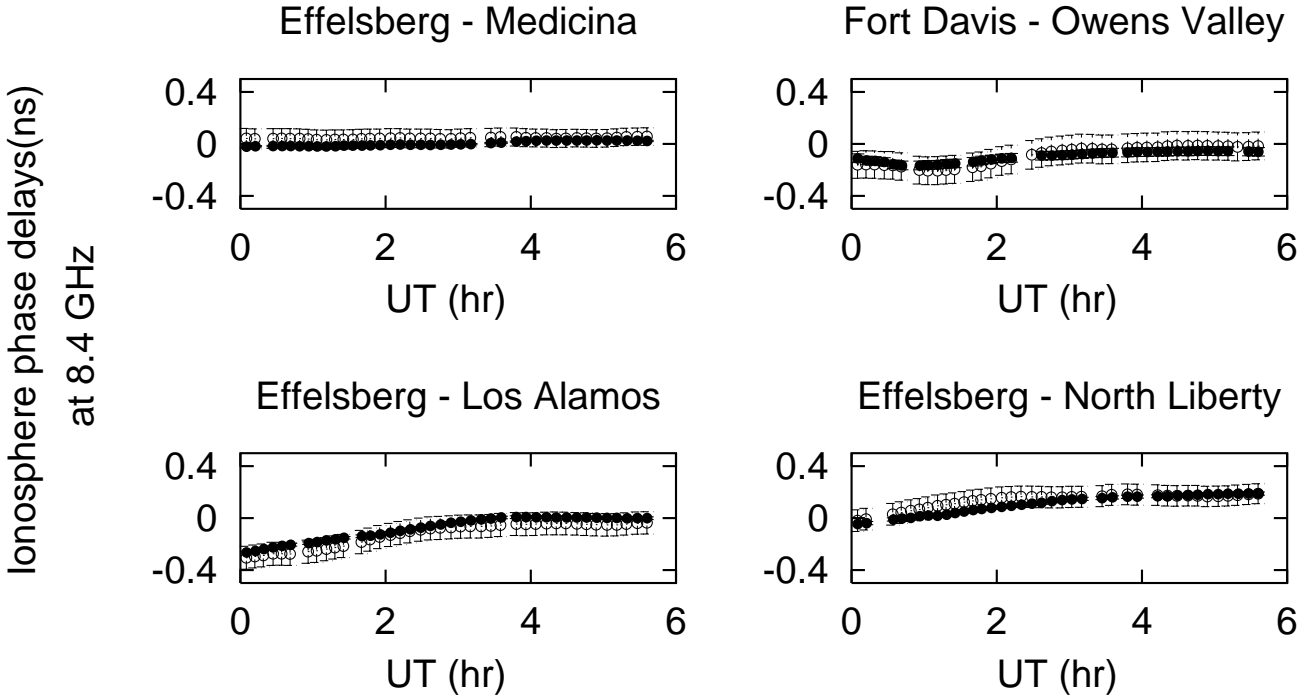}
\caption[]{
%\begin{small}
Ionosphere contribution, in nsec at 8.4 GHz, as obtained from our 
dual-frequency-band (8.4/2.3 GHz) measurements 
(filled circles) and from the GPS-based measurements
(open circles), for a
representative set of baselines including both intracontinental and 
intercontinental baselines.
The standard errors shown are for the TEC-inferred delays, and are much larger
than those from dual-frequency-band measurements.
All four plots start at 00:00,  since until approximately that 
time Effelsberg observed only at 8.4 GHz,  
and thus lacked dual-frequency-band corrections. 
%\end{small}
}
\label{fig:tec-plot}
\end{figure*}

As explained in Sect.~\ref{sec:intro},  
we removed most of the ionosphere contribution in two alternative ways, 
first by using our 
dual-frequency-band observations and, second,  
by using the total electron content (TEC) 
values deduced from GPS measurements 
(see, e.g., Sard\'on et al. \cite{sardon}). 
For the latter removal, 
we followed the same procedure as described in Ros et al. (2000),
modeling the ionosphere as a thin layer located at an
altitude of 350 km.
We used GPS data from Wettzell (Germany) to obtain TEC values for 
Effelsberg and Medicina, 
and from Goldstone (California) to obtain TEC values  
for Owens Valley, Los Alamos, and Fort Davis;
we were fortunate to have a GPS antenna collocated with our 
VLBI antenna at North Liberty to obtain TEC values for this site.
In Fig.~\ref{fig:tec-plot} we compare the estimated 
ionosphere delays as deduced from both dual-frequency-band observations
(8.4 GHz/2.3 GHz delays) and GPS-based measurements (TEC delays), 
for a representative subset of baselines, intracontinental 
(upper two plots) and intercontinental (lower two plots). 
Dual-frequency corrections refer only to baselines, not 
individual antennas.
On the other hand, TEC ionosphere corrections are calculated for individual
antennas.
Therefore, for each antenna there can be a constant offset 
between the TEC ionosphere correction and the dual-frequency one 
as a result of instrumental effects, but,  
for clarity, we subtracted the mean difference between each of the
two corrections
(Effelsberg, 1.35 ns; Medicina, 0.55 ns; North Liberty, 7.80 ns; 
Fort Davis and Los Alamos had no offset).
We assumed a statistical standard error for the vertical TEC at 
each GPS antenna of $2 \times 10^{16} \, {\rm m}^{-2}$ (Sard\'on,
priv. comm.),  
corresponding to a standard error
of $\sim \! 40$ ps for the (vertical) phase delays at 8.4 GHz.
For each antenna, 
this error is multiplied by the value from the mapping function 
for the elevation angle of each radio source
(here, the secant of the zenith angle at the ionospheric
point)
of each radio source. 
The resultant mapped TEC-based 
ionosphere corrections had standard errors
ranging from $\sim \! 70$ ps to $\sim \! 120$ ps,
highly correlated from point to point (Fig.~\ref{fig:tec-plot}). 
The standard errors for our dual-frequency-band corrections are the 
appropriate combination of the statistical errors of the phase
delays at each frequency, and ranged from $\sim \! 5$ ps for the
Effelsberg--Medicina baseline up to $\sim \! 30$ ps for 
the Fort Davis--Owens Valley baseline. 
The root-mean-square (rms) of the differences between
the corrections obtained by the two methods ranged from $\sim \! 20$ ps  
up to $\sim \! 90$ ps, depending on baseline. 
The maximum difference was for intercontinental baselines 
and was $\sim 110$ ps, when it was night in Europe and noon in North America.
The level of agreement in the results from the two methods is gratifying,
and foreshadows the agreement in the astrometric results obtained below with
the two methods.

\subsection{Opacity Contribution}
\label{sec:opacity}

When we use phase delays at both 8.4 and 2.3 GHz 
(see Sect.~\ref{sec:ionosphere}), 
the reference points on the 2.3 GHz maps 
should be chosen to correspond to the same points in the 
sky as those on the 8.4 GHz maps. 
Unfortunately, opacity effects 
may introduce an offset between the 
POB for one frequency
band with respect to the POB for the other frequency band 
(see, e.g., Marcaide \& Shapiro \cite{jmm84}).  
Since jet components are less self-absorbed than core components, 
their associated peaks should be less affected by opacity effects and 
should therefore be better suited for use as guides for the alignment of 
the maps at 8.4 and 2.3 GHz.
We registered the peaks of the less self-absorbed components in 
our 8.4 GHz maps with their corresponding peaks 
in (twofold overresolved) 2.3 GHz maps. 

For 1803+784, we identified component X3 at 8.4 GHz with component S3 at
2.3 GHz (Fig.~\ref{fig:1803}) on a twofold  overresolved map (not shown).
Based on this map, we concluded that blended components 
S1, S2, and S3 at 2.3 GHz correspond to 
components X1, X2, and X3, respectively, in the 8.4 GHz map. 
This identification implied that in our maps the POB
of 1803+784 at 2.3 GHz (S1) is offset by 
$\sim$ 0.6 mas westward from the POB at 8.4 GHz (X1). 
For 1150+812, we also used a twofold overresolved map at 2.3 GHz 
to make a plausible registration of the maps for 
the two bands.
We identified components X3 and X4 at 8.4 GHz with 
components S3 and S4 at 2.3 GHz (Fig.~\ref{fig:1150}),  
implying a shift of the POB at 2.3 GHz (S1) 
of $\sim$ 0.5 mas northward from 
the peak of the core in the 8.4 GHz map (X1).  
To estimate the standard error in our determination of those shifts 
we followed a similar procedure to that described for 
the determination of the standard error associated with the
reference-point identification (POB), but applied to 
components X3 and S3, for each source. 
This procedure yielded a standard error of 0.4 mas for each source. 
 Finally, to take into account the possible 
additional shifts due to opacity effects of the POB of component 
S3 with respect to X3, for each source, 
we increased the estimated standard error 
in each coordinate of the registration estimate from 0.4 mas to 0.7 mas. 
Estimates of the shift between the POBs 
at 8.4 and 2.3 GHz 
(where opacity effects are expected to be larger than in the 
jet components) of other radio sources -- 1038+528 A 
(Marcaide \& Shapiro~\cite{jmm84}), 1226+023 (Charlot~\cite{charlot93}),
1901+319 (Lara et al.~\cite{lara94}) -- are also no greater than 0.7 mas. 
Thus,  our estimates of these shifts 
for 1803+784 and 1150+812 seem reasonable.

In this dual-frequency-band method to reduce the ionosphere contribution, the 
resultant ``ionosphere-free'' phase delays have the form:

$$
\tau_{\phi , free} = \frac{R \, \tau_{\phi ,1} - \tau_{\phi ,2}}{R-1} 
$$
 
\noindent
where the subscripts 1 and 2 refer to the two frequency bands,
with $\nu_1 > \nu_2$ and $R=(\nu_1/\nu_2)^2$. 
%is the square ratio 
%between the center frequencies of the two bands -- 
In our case, $R\simeq (8.4/2.3)^2 \simeq 13$; 
hence standard errors in the phase delays at 8.4 GHz are scaled up
by a factor $R/(R-1) \simeq 1.08$, 
whereas corresponding errors in the phase
delays at 2.3 GHz are scaled down by $1/(R-1) \simeq 0.08$
in their effect on $\tau_{\phi , {\rm free}}$.
Likewise, uncertainties accounting for registration
errors due to opacity effects
have to be scaled by $1/(R-1)$.
Thus, the contribution to the total error budget
due to the use of dual-frequency-band measurements,
in particular the seemingly large error at 2.3 GHz discussed above,
is not significant (see entry ``Opacity effects'' in Table~\ref{tab:errors}).

\begin{table}[hbtp !]
%\begin{table*}[p]
\begin{flushleft}
\caption[]{Contributions to the standard errors of the estimates of the 
relative coordinates of the POB at 8.4 GHz of 1150+812 and 1803+784}
\label{tab:errors}
\begin{footnotesize}
\begin{tabular}{lll}
\hline
  Effect  & ${\sigma_{\Delta\alpha}}^{\rm a}$ 
          & ${\sigma_{\Delta\delta}}^{\rm a}$ 
\hspace*{0.015cm} \\
          &         mas            &      mas           \\
\hline
8.4 GHz source structure$^{\rm b}$     & 0.07 \hspace*{0.15cm} 
                                       & 0.07 \hspace*{0.03cm}\\
2.3 GHz source structure$^{\rm c}$     & 0.04 \hspace*{0.15cm} 
                                       & 0.04 \hspace*{0.03cm}\\
Opacity effects$^{\rm c}$              & 0.08 \hspace*{0.15cm} 
                                       & 0.08 \hspace*{0.03cm}\\
Statistical errors$^{\rm d}$           & 0.41 \hspace*{0.15cm} 
                                       & 0.62 \hspace*{0.03cm}\\
Software limitations$^{\rm e}$         & 0.40 \hspace*{0.15cm} 
                                       & 0.40 \hspace*{0.03cm}\\
\hline
Root-sum-square                        & 0.58 \hspace*{0.15cm} 
                                       & 0.75 \hspace*{0.03cm}\\
\hline
\end{tabular}
%\begin{scriptsize}
\end{footnotesize}
\noindent
%\begin{scriptsize}
\item[$^{\rm a}$] {Contributions to the standard errors of 
		  the estimates of the coordinates of 1803+784 
		  minus those of 1150+812, namely 
	          relative right ascension,  $\Delta\alpha$, 
	          and relative declination, $\Delta\delta$. 
                  } 
\item[$^{\rm b}$] {
Root-sum-square (rss) of the standard errors due to the use of 
data at 8.4 GHz. 
Each value has been scaled by 
$R/(R-1) \approx 1.08$. See Sect.~\ref{sec:opacity}. 
} 
\item[$^{\rm c}$]
{
rss of the
standard errors due to the use of data at 2.3 GHz, which includes
estimated errors in the ``registration'' of the 2.3 GHz and 8.4 GHz maps,
taken to be 0.7 mas for each source. 
Each value has been scaled by 
$1/(R-1) \approx 0.08$. See Sect.~\ref{sec:opacity}. 
} 
\item[$^{\rm d}$]
{
We estimated {$\sigma_{\Delta\alpha}$} in  milliseconds (ms). 
To transform this estimate into milliarcseconds (mas), 
we multiplied by $15 \cdot \cos 
\delta_{1803+784}$.
}
\item[$^{\rm e}$]
{See Sect.~\ref{sec:sens}}.
%\end{scriptsize}
\end{flushleft}
\end{table}

\section{Relative Position of the Two Sources}
\label{sec:relpos}

We obtained a final set of phase delays for each source 
by correcting for 
source-structure, opacity-effect, and ionosphere contributions
as described above.
We then formed a set of differenced
phase delays by subtracting the delay for each observation of 1803+784
from the corresponding delay for the previous observation of 1150+812.
The use of differenced phase delays 
is in general more effective the closer together the sources are in 
the sky, since differencing results from neighboring observations tends to
cancel effects which for each source alone cannot be accurately 
enough described by theoretical models. 
The best such pair of sources for such cancellation 
so far studied is 1038+528 A and B 
(Marcaide \& Shapiro \cite{jmm83}),  
because this pair simultaneously lies inside the beam of each antenna,
yielding almost complete cancellation of several sources of error.
For sources separated by increasingly larger angular 
distances, the cancellation of unmodeled effects in general lessens, 
due directly to the increase in angular separation of the sources and indirectly
to the increase of the cycle length of the observations.

\begin{table*}[htbp !]
\caption[]{Phase-delay-based estimates of the coordinates of 1803+784 minus those of 1150+812} 
\label{tab:positions}
\begin{flushleft}
\begin{footnotesize}
\begin{tabular}{lll}
\hline
  &  \multicolumn{2}{c}{Ionosphere correction method}  \\ \cline{2-3}
  & Dual-frequency {\rm $^a$}  & GPS-derived TEC {\rm $^b$} \\ 
  & \hspace*{15pt} (mas)  & \hspace*{15pt} (mas)   
\hspace*{0.015cm} \\
\hline
$\Delta\alpha - 6^h \, 7^m 33\fs 18473^{\rm c}$ 
           &  $-0.12 \pm 0.58^{\rm d}$ 
           &  $+0.12 \pm 0.78^{\rm d}$  \\
$\Delta\delta + 2^\circ \, 30' 25\farcs 13601^{\rm c}$ 
                        &   $-0.44 \pm 0.75$ 
			&   $-0.31 \pm 0.82$ \\
\hline
\end{tabular}
\end{footnotesize}
%\begin{scriptsize}
%\noindent
\item[$^{\rm a}$]{
  Estimates of the relative position of the two sources, 
  minus the reference separation, 
  from a weighted-least-squares analysis of the 
  phase delays, after corrections for 
  source-structure, source opacity, and ionosphere effects, 
  the last via dual-frequency-band observations. 
  The estimated standard  errors include statistical standard errors  
  and errors due to opacity effects, to 
  registration of the maps, and to software limitations. 
  See Table~\ref{tab:errors} and Sect.~\ref{sec:sens}.
  }
\item[$^{\rm b}$]{
  Estimates of the same relative position as above,
  except that only phase delays at 8.4 GHz were used and that
  the corrections for ionosphere effects were obtained from
  GPS-based estimates of TEC.
  }
\item[$^{\rm c}$]{
  The reference separations are from the IERS (\cite{iers}).
  See Table 1. 
  }
\item[$^{\rm d}$]{
To facilitate comparisons between $\Delta \alpha$ and $\Delta \delta$,
we have multiplied the $\Delta \alpha$ (ms) values by 
$15 \cdot \cos \delta_{1803+784}$ 
to obtain $\Delta \alpha$ in mas.
}
%\end{scriptsize}
\end{flushleft}
\end{table*}

\begin{figure*}[htb !]
\vspace*{100mm}
\includegraphics{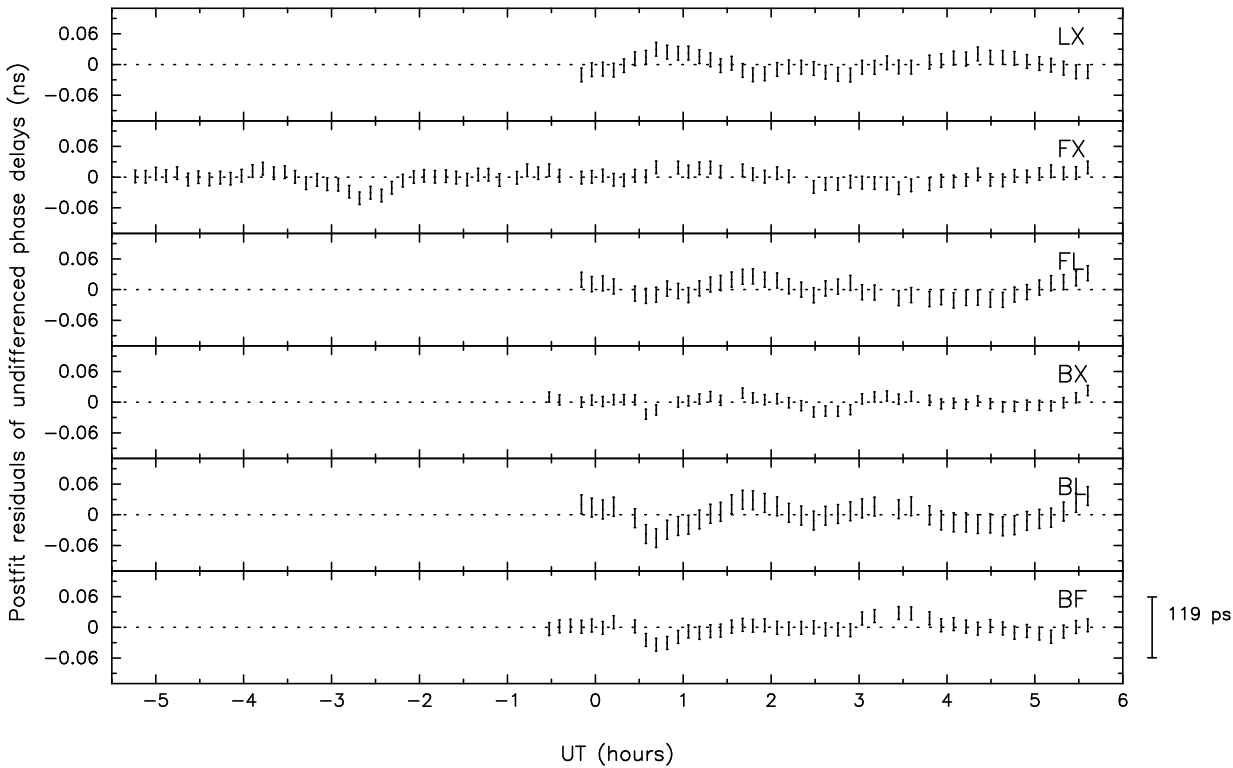}
\caption[]{
%\begin{scriptsize}
Postfit residuals of the phase delays (ns) 
of 1150+812 for the subset of 
baselines involving the following antennas: Effelsberg (B), 
Medicina (L), Fort Davis (F), and Los Alamos (X). 
The solution utilized corrections for ionosphere effects determined from 
our dual-frequency-band observations. For that 
reason, the baselines involving Effelsberg show a large time gap,
since this antenna observed in dual-frequency mode after 23:28.
Similarly, the Medicina antenna shows a gap until 23:48, since
the phase-delay connection for this antenna appeared uncertain
in this time range.
One ambiguity interval at 8.4 GHz is 119 ps
(shown at the right margin).
The error bars shown are the standard deviations of the 
phase delays, scaled so that the 
weighted-mean-square of the postfit residuals is unity 
for each baseline (see text).
%
%$\Chi^2$ per degree of freedom of the postfit residuals is unity
%
%\end{scriptsize}
}
\label{fig:res_undif}
\end{figure*}

\begin{figure*}[htb !]
%\vspace*{100mm}
\vspace*{300pt}
\includegraphics{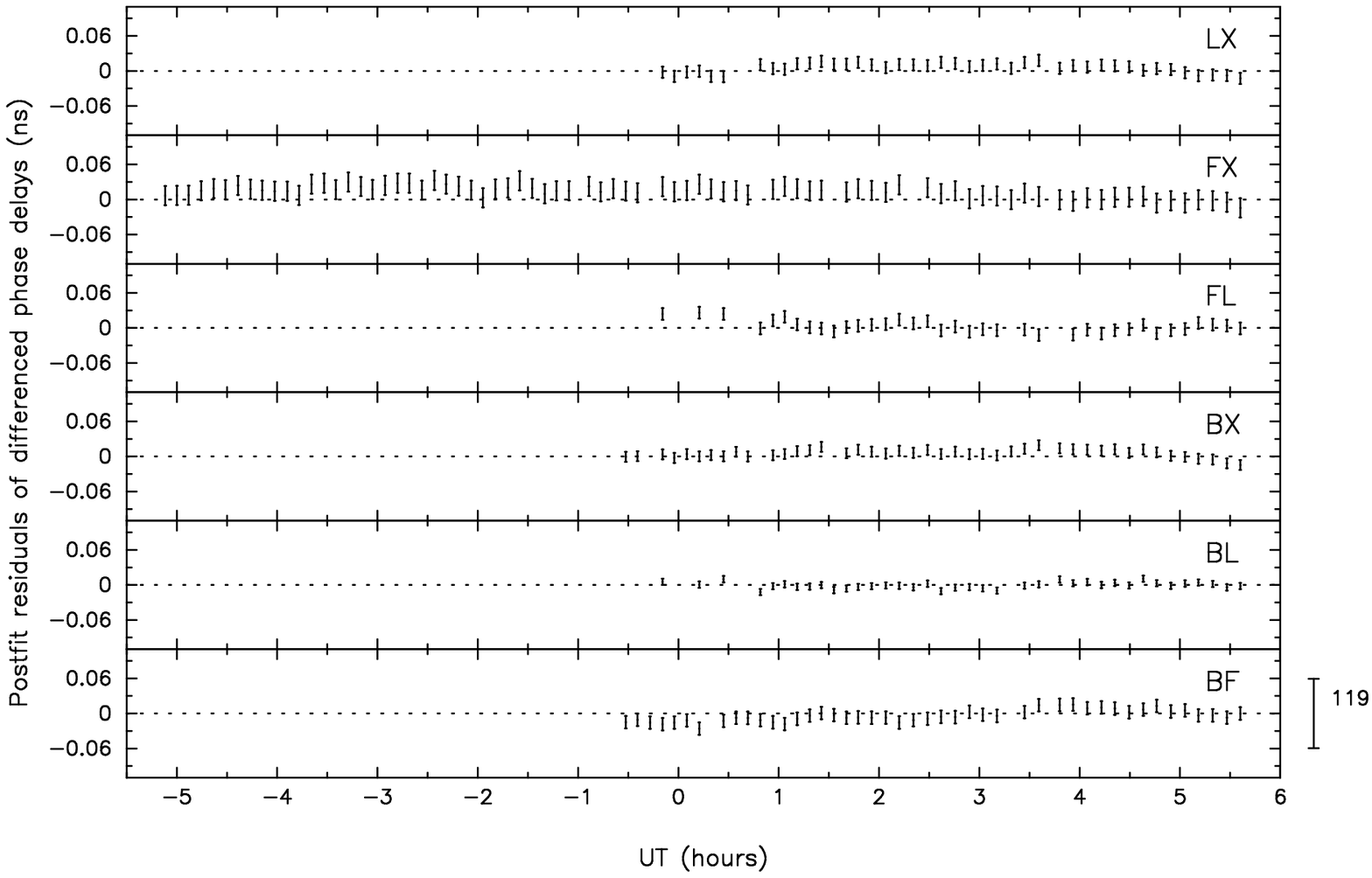}
\caption[]{
%\begin{scriptsize}
Postfit residuals of differenced phase delays (ns) for the same solution
and the same subset of 
baselines as in Fig.~\ref{fig:res_undif}. 
The error bars
shown are the standard deviations of the differenced phase delays, 
scaled so that the weighted-mean-square of the postfit residuals
is unity  for each baseline (see text).
%\end{scriptsize}
}
\label{fig:res_dif}
\end{figure*}

From a weighted-least-squares analysis 
of the differenced phase delays, 
we estimated the coordinates of 1803+784 relative to those 
of 1150+812 (see Table~\ref{tab:positions}). 
In this analysis, we also included the (undifferenced) 
phase delays of 1150+812, suitably weighted, 
to estimate the relative behavior of the site clocks.
(The opposite scheme, i.e., using the differenced phase delays
of 1803+784 minus those for 1150+812, and the undifferenced 
phase delays of 1803+784, did not alter our estimate significantly.)
The total number of parameters whose values were estimated was 26,
and include those of a third-degree polynomial to model the clock behavior 
at each station 
(except for North Liberty, which was taken as our reference station), 
the coordinates of 1803+784, 
and four atmosphere parameters (see Sect.~\ref{sec:troposphere})
for the Effelsberg antenna, since
the information on the atmosphere conditions 
provided for this site was very sparse, and thus
of doubtful utility. 
Nonetheless, the estimated atmosphere parameters for Effelsberg agreed with 
the a priori values within 0.11 ns.  

Our choice of 1150+812 as the reference source was motivated 
by its being one of the defining sources of the quasi-inertial 
International Celestial Reference Frame. 
As a test of the effect of this choice, 
we solved  for the coordinates of 1150+812--keeping 
those of 1803+784 fixed at their IERS values--and obtained  different 
results for 
$\Delta \alpha$ and $\Delta \delta$.
As a simple check on the significance of those differences -- one that
avoids the effects of the reference system in which the
coordinates are expressed -- we calculated 
the arclength between the two sources for each of the two solutions. 
We expected to obtain nearly the same result for the two cases and, 
indeed, 
the difference was a mere 5 $\mu{\rm as}$, negligible compared to the overall
standard errors of our results.

For our nominal, dual-frequency-band solution
described above, we scaled separately the standard deviations of 
the differenced phase delays 
and of the phase delays of 1803+784, so that for the data 
for each type, and for each baseline,
the weighted-mean-square of the postfit residuals was nearly unity.
As a result, the error bars assigned to the differenced phase delays 
were, in general, smaller by a factor $\sim$ 1.2 with respect to the 
error bars assigned to the phase delays of 1150+812
(see Figs.~\ref{fig:res_undif} and \ref{fig:res_dif}). 
We studied the effect of varying the relative weights of our two data 
types and verified that such variations did not change 
our result significantly.
In fact, any ratio of the weights of the differenced to the undifferenced 
data between 0.1 and 10 with respect to our ``nominal'' solution 
did not alter our estimate of the angular separation  significantly. 
Figs.~\ref{fig:res_undif} and ~\ref{fig:res_dif} 
also illustrate a virtue of the difference phase-delay technique
mentioned above:  
although the separation of the sources is 
$14^\circ\, 50'\, 21\farcs 147559$  
and the cycle length of the observations is correspondingly large, 
the differenced phase delays are partially free 
from some incompletely modeled effects as attested by 
the larger residuals seen in 
Fig.~\ref{fig:res_undif} than in Fig.~\ref{fig:res_dif},
in particular for the baselines involving the Medicina antenna.

The standard errors in Table~\ref{tab:positions} are 
larger than the differences between our estimates of $\Delta \alpha$
and $\Delta \delta$, and those provided by IERS, indicating
that the a priori IERS values are quite accurate. 
This accuracy is not surprising, since both radio sources are dominated 
by emission from the core, and so the IERS values -- mainly based on averages of 
results from observations 
at many epochs using group-delay astrometry -- should be accurate, 
except for a small offset that may be contributed by persistent opacity
effects  in the core components. 
For sources with strong emitting features well separated from their cores,
larger differences from the a priori IERS values can be expected, 
given the greater sensitivity of the ``structure group delay'' to components
farther from the cores (Thomas~\cite{thomas80}).
Those differences can be largely removed by using the more precise,  
structure-free phase delays.

\section{Sensitivity Analysis}
\label{sec:sens}

We carried out a sensitivity analysis to estimate the 
contributions of individual effects 
to the standard errors in the relative-position determination; 
see Table~\ref{tab:sens} for a list of these effects.
We found that the primary source of error is due to 
the uncertainty in our knowledge of  
the neutral atmosphere, 
which by itself contributes 
a standard error of about 0.4 mas in right ascension and about 
0.6 mas in declination to the estimate of the position of 1803+784.
To be conservative in our estimate of the 
error contributed by the neutral atmosphere, 
we took for each site the larger of the following two values: 
the root-sum-square of the delay contribution
from all atmosphere parameters for that site (one every three hours), 
or the algebraic sum of the signed partials.
In this way, we made some allowance for the potential effects of 
possible correlations among the estimates of the atmosphere parameters
for each site.
The second largest contribution to the total error
budget of each coordinate comes from 
the  corresponding standard error in the a priori position used for 
the reference source (1150+812).
This large contribution is 
due to the geometry of the sources on the sky.
For sources with smaller angular separations,
the contribution of the a priori uncertainties of the coordinates
of the reference source to the standard error 
in relative position is usually a small 
fraction of the a priori uncertainty of the reference coordinates 
(e.g., Guirado et al.~\cite{gui95b}),  
because there is a positive correlation 
between the right ascensions, and similarly the declinations, of the two sources.
However, because 1150+812 and 1803+784 are 
within $12 \degr$ of the north celestial pole and lie almost exactly 
six hours apart in right ascension, 
an error in the position of one of the sources in right ascension
(or declination) translates into a similar error  for 
the other source in declination (or right ascension).
(See Table~\ref{tab:sens}.) 
A covariance analysis confirms this ``tradeoff'', showing 
a fairly strong correlation between the 
estimates of the right ascension of one source and 
the declination of the other source
($\rho (\alpha_{1150+812}, \delta_{1803+784}) \approx -0.7;  
\rho (\delta_{1150+812}, \alpha_{1803+784}) \approx 0.7$),
but little correlation between the estimates of the two right 
ascensions
(and between the two declinations).
Moreover, the nearly complete cancellation of the geometric errors usually obtained in 
astrometric studies of radio sources with smaller angular separations
is not so effective in the present case.
For example, the uncertainties in UT1--UTC values do contribute 
significantly to the 
standard error in the estimate of the angular separation.
The remaining significant contributions 
to the standard error are attributable to 
the uncertainties in the estimates of the antenna locations. 

\begin{table}[htb !]
%\begin{table}[p]
\caption[]{Sensitivity of the estimated position of 1803+784 to 
the values of other model parameters included in 
the weighted-least-squares analysis.}
\begin{flushleft}
\begin{footnotesize}
\begin{tabular}{llrrr}
\hline
  Parameter  & & Standard & {$\delta \Delta\alpha$} 
			  & $\delta \Delta\delta$ 
\hspace*{0.03cm} \\
            & & Deviation$^{\rm a}$        & ($\mu$as)$^{\rm b}$ &        ($\mu$as) \\
\hline
${\alpha_{1150+812}}^{\rm c}$ & & 174 $\mu$as  &  --6  \hspace*{0.15cm} &  --158 \hspace*{0.03cm}\\
${\delta_{1150+812}}^{\rm c}$ & & 140 $\mu$as  & 135  \hspace*{0.15cm} &  --2 \hspace*{0.03cm}\\
%      B & 0.17 ns/3 hr  & --162 \hspace*{0.15cm}    &  99  \hspace*{0.03cm}\\
Atmosphere   &  
    F & 0.17 ns/3 hr     & --123 \hspace*{0.15cm} & 298  \hspace*{0.03cm}\\
zenith delays$^{\rm d}$:    & 
       I & 0.17 ns/3 hr     & --150 \hspace*{0.15cm} & --87   \hspace*{0.03cm}\\
      & L & 0.17 ns/3 hr     & --267 \hspace*{0.15cm} & 117 \hspace*{0.03cm}\\
      & O & 0.17 ns/3 hr     &  111 \hspace*{0.15cm} & 415 \hspace*{0.03cm}\\
      & X & 0.17 ns/3 hr     & --45  \hspace*{0.15cm} & 227 \hspace*{0.03cm}\\
Site coordinates$^{\rm e}$: & B & 2 cm       & 60 \hspace*{0.15cm} & 40   \hspace*{0.03cm}\\
                  & F & 2 cm       & 30 \hspace*{0.15cm} & 35 \hspace*{0.03cm}\\
                  & I & 2 cm       & 42 \hspace*{0.15cm} & 20 \hspace*{0.03cm}\\
                  & L & 2 cm       & 69 \hspace*{0.15cm} & 32 \hspace*{0.03cm}\\
                  & O & 2 cm       & 34 \hspace*{0.15cm} & 76 \hspace*{0.03cm}\\
                  & X & 2 cm       & 24 \hspace*{0.15cm} & 39 \hspace*{0.03cm}\\
Earth's pole$^{\rm c}$:     & x & 0.7 mas    & 18  \hspace*{0.15cm} & --1 \hspace*{0.03cm}\\
                  & y & 0.7 mas    &     8 \hspace*{0.15cm}&  21  \hspace*{0.03cm}\\
UT1--UTC$^{\rm c}$           &   & 0.04 ms   &  -126 \hspace*{0.15cm}& -95 \hspace*{0.03cm}\\
\hline
\hline
Root-Sum-Square   &   &            & {\bf 412} \hspace*{0.15cm}
				   & {\bf 616} \hspace*{0.03cm}\\ 
\hline
\label{tab:sens}
\end{tabular}
\end{footnotesize}
%\begin{scriptsize}
\noindent
\item[$^{\rm a}$]
{Each value in the $\delta \Delta\alpha$ and $\delta \Delta\delta$
columns was obtained by computing the effect 
of altering the a priori value of each parameter
by one standard deviation on the relative-position determination. 
%The resultant entry is the value of 
%the difference between each result and the ``nominal'' value. 
}
\item[$^{\rm b}$]
{
We have multiplied $\mu$s by $15 \cdot \cos \delta_{1803+784}$ 
to obtain $\mu$as for the penultimate column.
}
\item[$^{\rm c}$]
{
The standard deviations are as provided in IERS (\cite{iers}). 
For the sensitivity analysis, we adopted  
standard deviations 
for the coordinates of 1150+812  twofold  larger than
the quoted IERS (\cite{iers}) errors.
%Our tabulated sensitivities 
%can be used to reestimate standard errors in  
%any of the parameters listed in the table
%by appropriately  scaling those standard errors.
}
\item[$^{\rm d}$]
{
We specified the atmosphere delays at each site by a 
piecewise-linear function characterized by values at epochs about
three hours apart. 
For each such epoch, we used 0.17 ns as the standard error. 
The quoted sensitivity values take into account the 
total contribution, for each antenna, from all atmosphere parameters.
Effelsberg (B) is not included, since we solved for its
atmosphere parameters. 
(See Sect.~\ref{sec:troposphere}.)
}
\item[$^{\rm e}$]
{
As given in Table 1. The 2 cm standard deviation 
refers to each of the three coordinates for each antenna site.
The sensitivity values account for the errors in all three coordinates,
assumed to be independent.
}
%\end{scriptsize}
\end{flushleft}
\end{table}

Software limitations prevented us from taking into account
in our analysis such effects as ocean and atmosphere loading, 
tidal terms in polar motion and UT1, and arbitrary variations in the
atmosphere zenith delays.
Based upon a surrogate test of the sensitivity of the relative
position of a different pair of radio sources 
(observed in an unrelated experiment) to these effects, 
performed using the CALC (Ma et al.~\cite{ma86}) 
and SOLVK (Herring et al.~\cite{herring90}) 
packages, we estimated that the known limitations of our software 
contribute an uncertainty of about 0.4 mas to each coordinate of our
estimate of the sources' relative position.

%%%%%%%%%%%%%%%%%%RESULTS AND DISCUSSION %%%%%%%%%%%%%%%%%%%%%%%%%%%%%%%

\section{Summary}
\label{sec:results}

We observed the strong radio sources 1150+812 and 1803+784 
with a VLBI array on 18 November 1993, an epoch of mild solar activity. 
The antennas recorded data simultaneously at 8.4 and 2.3 GHz, 
which allowed us to estimate the ionosphere contributions to the 
phase delays; we also used TEC values from GPS measurements to
estimate such contributions.
We estimated, and thereby partly removed from the phase-delay data,  
contributions due to the dry and wet components of the atmosphere, 
and the brightness structure of the sources. 
The phase-connection process, required to extract the precision inherent in 
the phase delays, did not pose special difficulties 
for our 7 min cycle time,
indicating that spatial and temporal fluctuations due to the atmosphere
and the ionosphere were not large enough to prevent reliable phase
connection in either frequency band.

We then estimated the relative position of the sources 
(Table~\ref{tab:positions}) via a weighted-least-squares
analysis of a combined data set of undifferenced and differenced phase delays. 
The estimates resulting from use of GPS-based TEC values to account 
for ionosphere effects agree  
with those obtained from our dual-frequency-band VLBI measurements 
to within the standard errors for each method.
Ros et al. (\cite{ros99},~\cite{ros2000}) also successfully used GPS-based
TEC values to remove the ionosphere's contribution to the phase delays
at 8.4 GHz. 
The checking of their removal in that experiment was limited to the precision of 
the dual-frequency-band  (8.4 and 2.3 GHz) group delays, since 
Ros et al. could not reliably connect their phase delays at 2.3 GHz. 
In our case, we successfully connected the phase delays at both 
frequencies, and thus for the first time 
compared TEC ionosphere delays against phase-delay-based ionosphere corrections
(Fig.~\ref{fig:tec-plot}).  
This agreement supports the use of single-band ($\geq 8.4$ GHz) 
VLBI observations 
for astrometry purposes, an option of particular interest for experiments 
to be carried out at epochs of strong solar activity, 
when phase connection for a frequency band $\leq 2.3$ GHz is likely to fail.  
Single-frequency VLBI observations along with GPS-derived TEC ionosphere
corrections can be confidently used instead.

Our standard errors shown in Table~\ref{tab:positions} are 
larger than the difference between our estimates 
of relative source position and those provided by IERS. 
The high level of agreement is likely due to the 
fact that both radio sources are strongly core-dominated.
However, were a bright emitting feature present several milliarcseconds
from one of the cores, large offsets could exist between the true core position
and the IERS position.
With VLBI phase-delay astrometry, such errors can be effectively 
reduced using structure-phase corrections computed from self-calibrated
maps derived from the same VLBI observations.
Moreover, phase-delay astrometry has the advantage of allowing us
to know {\sl very} accurately to which feature {\sl inside} each source 
we are referring when calculating relative positions.

In summary,
we have determined with submilliarcsecond accuracy the angular separation 
of two radio sources separated by almost $15\degr$, 
using phase delays from dual-frequency-band VLBI measurements. 
Since within $15\degr$ of any given source there are 
almost always two or more reasonably strong radio sources, 
we have thus demonstrated that phase delays should be usable for 
full-sky astrometry of radio sources. 
We have also shown that GPS-based measurements can be used to 
obtain reliable ionosphere corrections to the phase delays,
thus demonstrating the feasibility of conducting 
VLBI solely observations at frequencies $\geq 8.4$ GHz for astrometric purposes.
Future improvements in the modeling of the atmospheric delay contribution,
and in the knowledge of Earth rotation 
and pole position, as well as antenna location, 
should result in increasingly accurate estimates of the 
relative positions of sources far apart on the sky.

We stress that phase delays  are more reliably corrected for structure
phase than are group delays. 
Therefore, differenced phase-delay astrometry is better suited than 
group-delay astrometry for carrying out astrometric studies of extended radio sources. 
To date, however, group-delay astrometry has been used to establish 
a quasi-inertial celestial reference frame based on 
estimates of the positions of a number of relatively 
compact extragalactic radio sources from many years of regular observations.
Our results open the avenue to an alternative, potentially more 
accurate, approach, namely that of carrying out 
a suitable series of observing sessions and using 
difference phase-delay astrometry to obtain
submilliarcsecond positions for the cores of these sources.

\begin{acknowledgements}
We are grateful to the referee, Jim Ulvestad, for valuable comments and
suggestions.
We thank the staffs
of all the observatories for their contribution to the observations, and
in particular the MPIfR staff for their efforts during the correlation.
We also thank Dan Lebach for his work with the CALC/SOLVK package to test 
the quantitative significance of the limitations of our VLBI3 program.
This work has been partially supported by
the Spanish DGICYT grants PB93-0030 and PB96-0782, and by the European 
Comission's TMR-LSF program under contract No. ERBFMGECT950012. 
The VLA is a facility of the National Radio Astronomy Observatory. 
The NRAO is a facility of the National Science Foundation operated under 
cooperative agreement by Associated Universities, Inc.
\end{acknowledgements}

\end{document}